\documentclass[superscriptaddress,showkeys,twocolumn,nofootinbib,longbibliography]{revtex4-1}

\usepackage{amsmath}
\usepackage{amssymb}
\usepackage{graphicx}
\usepackage{epsf}
\usepackage{slashed}
\usepackage{enumitem}
\usepackage[czech,english]{babel}
\usepackage[cp1250]{inputenc}
\usepackage{hyperref}
\usepackage{xcolor}
\usepackage{physics}

\usepackage[only,llbracket,rrbracket,llparenthesis,rrparenthesis]{stmaryrd} 
\usepackage{accsupp} 

\usepackage[font=small,labelfont=bf,justification=raggedright]{caption}
\usepackage{subcaption}

\usepackage{bbm} 

\usepackage[nodayofweek]{datetime}
\newdateformat{mydate}{\twodigit{\THEDAY}{ }\shortmonthname[\THEMONTH], \THEYEAR}
\usepackage[normalem]{ulem}

\usepackage{color}
\usepackage{ulem} 

\definecolor{myred}{rgb}{1,0,0}
\definecolor{mygreen}{rgb}{0,0.8,0.2}
\definecolor{myblue}{rgb}{0,0,1}

\definecolor{Ared}{rgb}{1,0.7,0}
\definecolor{Agreen}{rgb}{0.7,0.8,0.2}
\definecolor{Ablue}{rgb}{0,0.7,1}

\renewcommand{\emph}[1]{\textit{#1}}

\begin{document}
\title{Unified theory of oscillons and modes}

\author{F. Blaschke}

\affiliation{Research Center for Theoretical Physics and Astrophysics, Institute of Physics, Silesian University in Opava, Bezrucovo
namesti 1150/13, Opava, 746-01, Czech Republic}
\affiliation{Institute of Experimental and Applied Physics, Czech Technical University in Prague, Husova 240/5, Prague 1, 110-00, Czech Republic}

\author{T. Romanczukiewicz}

\affiliation{Institute of Theoretical Physics, Jagiellonian University,
Lojasiewicza 11, Krak\'{o}w, Poland}

\author{K. Slawinska}

\affiliation{Institute of Theoretical Physics, Jagiellonian University,
Lojasiewicza 11, Krak\'{o}w, Poland}

\author{A. Wereszczynski}

\affiliation{Institute of Theoretical Physics, Jagiellonian University,
Lojasiewicza 11, Krak\'{o}w, Poland}
\affiliation{International Institute for Sustainability with Knotted Chiral Meta Matter (WPI-SKCM$^{\; 2}$), Hiroshima University, 1-3-1 Kagamiyama, Higashi-Hiroshima, Hiroshima 739-8531, JAPAN}

\begin{abstract}
We show that a (1+1) dimensional oscillon can be understood as a localized  resonance, i.e., a discrete, non-normalizable mode. Specifically, oscillon in the vacuum arises from the threshold mode of the corresponding Klein-Gordon equation, which, because of nonlinearity, gets localized. Following this idea, we find {\it wobblerons} - nonlinear excitations of kinks, that is, oscillons-kink bound state. This originates in the threshold or in an antibound mode. 
\end{abstract}

\maketitle

\section{Introduction}

Oscillons \cite{Bogolyubsky:1976nx, Gleiser:1993pt, Copeland:1995fq} are the most common soliton-like solutions existing in many non-linear field theories. They are coherent excitations that probe the non-perturbative sector also in the quantum regime \cite{Evslin:2025hjt}. They are an almost localized, quasi-periodic oscillation of the field where a small but continuous flux of radiation is emitted all the time. Therefore, they are not ultimately stable. Nevertheless, they can live for an extremely long time \cite{Honda:2001xg, Arodz:2007jh,Saffin:2006yk, Gleiser:2009ys,Fodor:2009kf, Salmi:2012ta,Saffin:2016kof, Olle:2020qqy, Zhou:2024mea, Li:2025ioq}, which makes them phenomenologically interesting, e.g., in condense matter \cite{2015OptSp.119..363R, 2015PhRvA..91b3631S, DiMauroVillari:2019mdb}. In cosmological context, they are expected to be produced in the early universe 
\cite{Amin:2011hj, Zhou:2013tsa, Lozanov:2014zfa, Olle:2019kbo,Iarygina:2020dwe, Hiramatsu:2020obh,Kawasaki:2020jnw, Mahbub:2023faw, Aurrekoetxea:2023jwd} or during collisions of cosmic defects \cite{Hindmarsh:2006ur, Gleiser:2007te}. They also exist in extensions of the Standard Model \cite{Graham:2006vy, Farhi:2005rz, Drogosz:2026nuh}.

Despite many applications, the origin of their existence is still unclear. In contrast to topological solitons and $Q$-balls \cite{Friedberg:1976me, Coleman:1985ki}, it is not related to any conserved topological or Noether charge. A non-zero mass threshold, i.e., a non-zero mass of the small perturbations in the vacuum, and flatter-than-quadratic behaviour of the potential near the vacuum were pointed out as conditions that guarantee the appearance of oscillons \cite{Gleiser:2009ys, Fodor:2008es}. However, some counter-examples are known \cite{Dorey:2023sjh, vanDissel:2025xqn, Martinez:2026hki}. 

Even such a characteristic property of oscillons as amplitude modulation is not fully explained. The most popular approximate description of oscillons does not take them into account at all \cite{Fodor:2008es}. Recently, it has been suggested that such modulations can be an effect of the formation of a two-oscillon bound state, where each unmodulated oscillon enters with its own fundamental frequency \cite{Blaschke:2024uec}. For us, it is important to note that the second oscillon is taken from the mass threshold. This idea has been further refined in the renormalization group-inspired framework \cite{Blaschke:2024dlt}.

There are other quasi-periodic excitations that are common in non-linear dynamical systems. They are bound (oscillating) modes that usually exist on solitons. They are linearly stable but decay in a power-law manner via non-linear interactions.

In the present work, contrary to the current state of
art, we show that nonlinear oscillons and linear modes
are very closely related to each other. We claim
that the oscillons are seeded by linear, discrete, but
non-normalizable, modes. For oscillons living in
vacuum, it is the threshold mode, while for oscillons living
in the presence of a kink, it can be a resonant mode,
or, speaking precisely, a non-normalizable antibound mode, with the frequency still below the mass threshold. 

Such an oscillon on top of the kink can be viewed as an example of a non-linearly excited kink, which we will call a
{\it wobbleron} - a non-integrable counterpart of the
sine-Gordon wobbler.

The main result can be reformulated as a unification
of oscillons and modes, where the oscillon is a nonlinear bound state that arises from threshold or antibound
resonance, which, due to the non-linearity of the model,
gets localized.
\section{Oscillons in the kink background}
Let us first stress that the Klein-Gordon equation in (1+1) dimensions 
\begin{equation}
    \phi_{tt}-\phi_{xx}+m^2\phi=0
\end{equation}
supports a non-normalizable threshold mode $\phi=Ae^{imt}$, provided that the mass of small fluctuations $m$ is not zero. This is a spatially-independent mode that uniformly oscillates in the entire space. 

The first evidence for our hypothesis that such a non-normalizable resonance triggers oscillons comes from the well-established perturbative expansion of a small-amplitude oscillon \cite{Fodor:2008es} or from the renormalization group framework \cite{Blaschke:2024dlt}. In both approaches, the first (linear) order solution is the threshold mode with frequency $\omega=m$ (or, in full generality, a massive plane wave) whose amplitude becomes an exponentially localized function due to non-linear corrections arising in higher orders. The non-linearities move the frequency below the mass threshold, $\omega_{osc}<m$, see \cite{Fodor:2008es} and \cite{Blaschke:2024dlt} for details. 

We want to exploit this point of view by considering the field $\phi$ not in vacuum (oscillon in vacuum) but in a non-trivial background given, e.g., by the kink. For that, we introduce a generic scalar field theory 
\begin{equation}
    \mathcal{L}=\frac{1}{2} \bigl(\partial_\mu \phi \bigr)^2 - U(\phi),
\end{equation}
which amounts to a non-linear Klein-Gordon equation
$\phi_{tt}-\phi_{xx}+\frac{dU}{d\phi}=0$. The potential $U$ is assumed to support a static kink solution $\phi_K$. Now, we expand the field into the (static) kink and a time-dependent (oscillating) component 
\begin{equation}
    \phi(x,t)=\phi_K(x)+\psi(x,t), \;\;\; \psi(x,t)=\psi(x)\cos (\omega t)
\end{equation}
where, as in \cite{Fodor:2008es, Blaschke:2024dlt}, small amplitude solutions are sought via the expansion 
\begin{equation}
    \psi=\sum_{n=1}^\infty \epsilon^n \psi_n, \;\;\;\;\; \omega^2= m^2+ \sum_{n=1}^\infty \epsilon^n \omega_n,
\end{equation}
with $\epsilon$ being a small parameter related to the amplitude of the oscillon. It is also convenient to use rescaled coordinates $\zeta=\epsilon x, \tau=\omega(\epsilon) t$. Then we arrive at
\begin{equation}
    -\omega^2 \ddot{\psi} + \epsilon^2 \mathbb{L} \psi = m^2\psi +\sum_{k\geq 2} g_k(x) \psi^k,
\end{equation}
where $\mathbb{L}=\frac{d^2}{d\zeta^2}- \frac{1}{\epsilon^2} V\bigl(\zeta/\epsilon\bigl)$. Next,
$V(x)= U^{''} (\phi_K)-m^2,$ is the background kink potential and $g_k(x)=\frac{1}{k!} U^{(k+1)}(\phi_K)$ are higher nonlinear corrections. 

As usual, at the linear order we find a massive Klein-Gordon equation
\begin{equation}
    \omega^2 \ddot{\psi}_1+m^2\psi_1=0
\end{equation}
with the threshold frequency solution, $\omega=m$, whose spatial behaviour is not specified at this order of the expansion
\begin{equation}
    \psi_1(\zeta,\tau)=p_1(\zeta) \cos \tau. 
\end{equation}
It is widely known that such a perturbative expansion produces unphysical, divergent secular resonant terms. They have to be removed. In $\epsilon^2$ order, it leads to the condition $\omega_1=0$. At $\epsilon^3$ order, equation for $\psi_3$ contains the resonant part which is proportional to 
$\left(\mathbb{L}p_1+\omega_2p_1+\lambda(x) p_1^3 \right)\cos \tau$, 
where now, contrary to the vacuum case \cite{Fodor:2008es}, $\lambda(x)$ is a kink-background-depending function
\begin{equation}
    \lambda(x)=\frac{5}{6 m^2} g^2_2(x)-\frac{3}{4}g_3(x).
\end{equation}
Requiring the absence of the resonant term, we find the following equation for the spatial profile $p_1$
\begin{equation}
    \mathbb{L}p_1 (\zeta)+\omega_2p_1(\zeta)+\lambda\left(\frac{\zeta}{\epsilon}\right) p_1^3(\zeta)=0.
\end{equation}
This is the leading-order profile of the oscillon in the background of the kink, i.e., the wobbleron. The wobblerons can be interpreted as non-linear bound states bifurcating from  resonances, where the 
$\mathcal{O}(\epsilon^3)$ correction makes its profile $L^2$ integrable.

\section{Localization of the resonance}
Let us closer focus on the relation to the threshold or anti-bound mode $\eta$ and rewrite the equation for the leading-order profile $p_1(x)$ (in the original coordinate $x$) 
\begin{equation}
-\mathcal{H} p_1 + \epsilon^{2}\Bigl(\omega_2 p_1 + \lambda(x)p_1^3\Bigr)=0, \label{full}
\end{equation}
where $\mathcal{H}=-\partial_x^2+V(x)$ is the linearized operator around the kink. At $\epsilon=0$ this reduces to
\begin{equation}
\mathcal{H} \eta = - \kappa^2 \eta,
\end{equation}
whose solutions correspond, beside the usual {\it bound modes}, to {\it threshold} ($\kappa^2=0$) and{\it anti-bound modes} ($\kappa^2>0$). In particular, for non-normalizable resonances, the solution satisfies
\begin{equation}
\eta(x) \to \mathrm{const} \cdot  e^{\kappa |x|} \quad \text{as} \quad |x|\to\infty,
\end{equation}
and is therefore not square-integrable.
As a consequence, the standard Fredholm solvability condition, based on the $L^2$ inner product, is not directly applicable. 

This issue is resolved by noting that the nonlinear solution is not given by the bare non-normalizable mode, but instead acquires a slowly varying envelope on the scale $x\sim \epsilon^{-1}$. The leading-order profile can therefore be written as
\begin{equation}
p_1(x) \sim \eta(x)\,F(\epsilon x), \label{p1}
\end{equation}
where $F$ should be a localized function. 

Acting with Hamiltonian $\mathcal{H}$ on (\ref{p1}) and expanding derivatives
we find that the following asymptotic form of the oscillon profile (for details see Appendix \ref{app:envelope})
\begin{equation}
p_1(x) \sim \eta(x)\,\mathrm{sech}(\epsilon x).
\label{approx_p1}
\end{equation}
It demonstrates that oscillons can be interpreted as nonlinear bound states arising from resonances via slow spatial modulation.
\section{Wobbelron}
\begin{figure*}
	\includegraphics[width=\columnwidth]{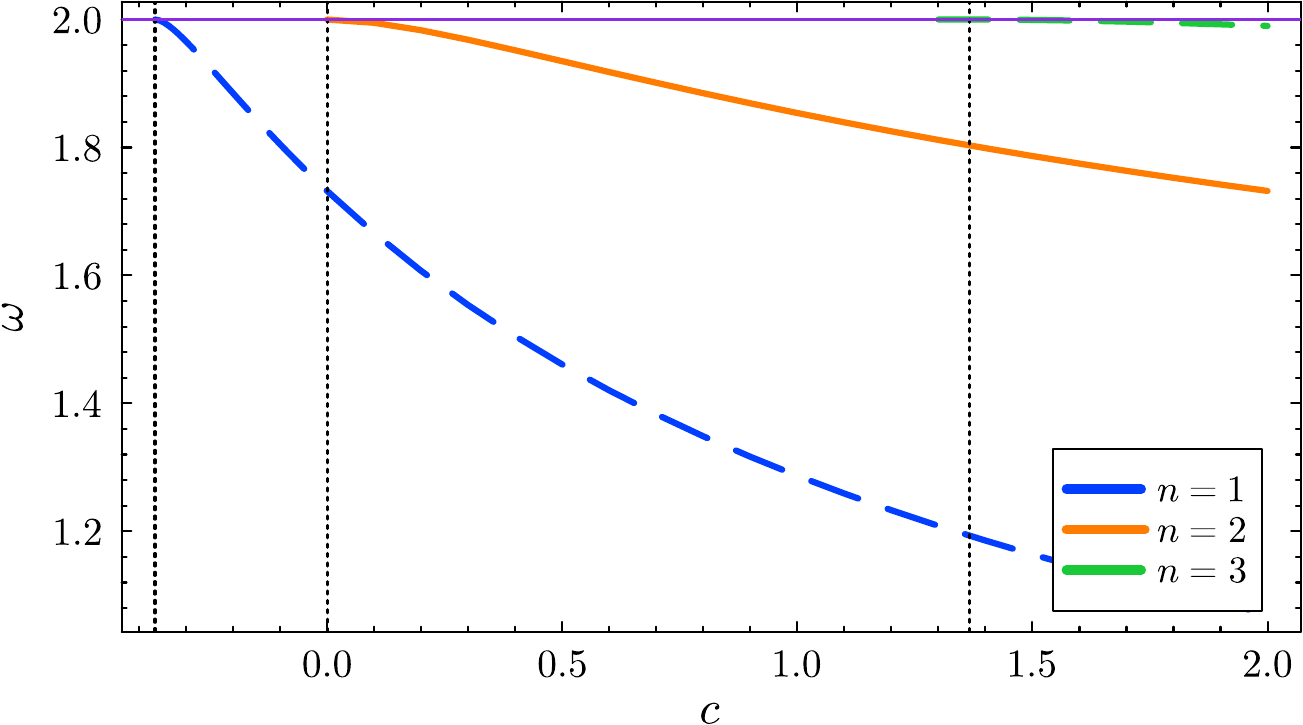} 
    \includegraphics[width=\columnwidth]{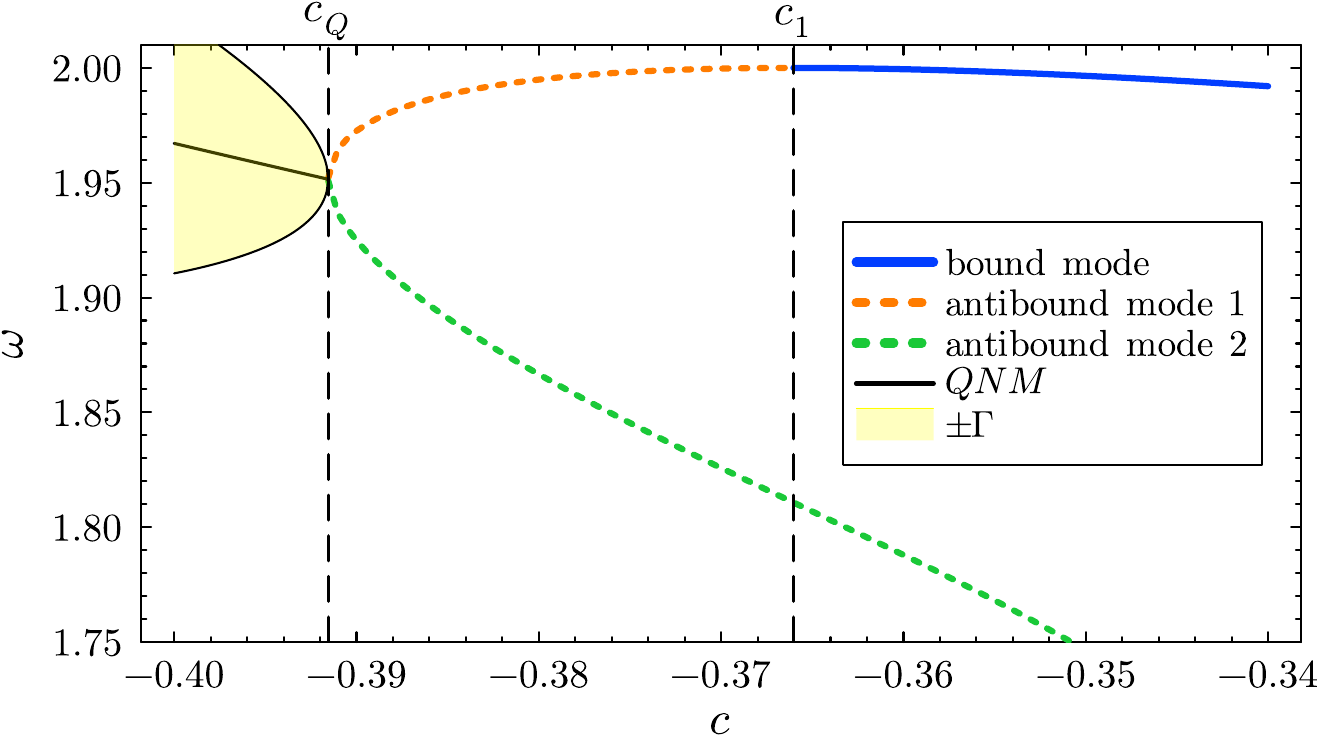} 
	\caption{Left: spectral structure of the CL theory. For all $c$, there is also a zero mode. The vertical dotted lines denote the positions of $c_1,c_2, c_3$ where pertinent threshold modes appear. Right: zoomed version in the neighborhood of $c=c_1$.}
    \label{CL-spectrum}
\end{figure*}

The first widely known example of a wobbleron is the wobbler solution in the sine-Gordon (sG) model. Due to integrability, it is a stable solution which can be written as a sum of the sG kink and an oscillating component $\phi(x,t)=\phi_K(x)+ \psi(x,t; \omega)$. Importantly, expanding the oscillating part at $\omega \to 1$ we find (see Appendix \ref{app:sG})
\begin{equation}
    \psi(x,t)= 4\epsilon \sin (\omega t) \frac{\tanh x}{\cosh (\epsilon x)}, 
\end{equation}
where $\epsilon=\sqrt{1-\omega^2}$. This is nothing but the threshold mode $\eta_{th}=\tanh x \sin t$ of the sG kink, localized by the envelope $F(\epsilon x)=\sech (\epsilon x)$ and with the reduced frequency $\omega <1$, exactly as predicted above. 

As a generic non-integrable example, we will analyze the Christ-Lee (CL) theory, which is defined by the following one-parameter family of potentials
\begin{equation}
    U_{CL} = \frac{1}{2} \frac{1+c\phi^2}{1+c}\bigl(\phi^2-1\bigr)^2.
\end{equation}
Here, the parameter $c>-1$. For $c=0$ we find the $\phi^4$ model, while for $c\to \infty$ we arrive at the $\phi^6$ model. The mass threshold is $m=2$.

In Fig.  \ref{CL-spectrum}, left panel, we present the structure of the linear discrete modes in a function of the parameter $c$. In particular, for $c<c_1=-0.36603$ there is no bound mode except the zero mode. For $c=c_1$ an odd threshold mode appears, which for $c>c_1$ becomes an odd bound mode. Then, for $c=c_2=0$ a new, even threshold mode shows up. Again, for $c>c_2$ it goes below the mass threshold and becomes an even (second) bound mode. Finally, for $c_3=1.36603$ another odd threshold mode appears. This repeats as $c$ further grows leading to  more and more complicated mode structure. From our point of view, the regime close to $c=0$ is especially interesting.

Importantly, the bound mode, after becoming the threshold mode, does not disappear, see Fig. \ref{CL-spectrum}, right panel, where we show the spectrum close $c=c_1$. In contrast, it transmutes to an anti-bound mode. 

The anti-bound mode at some smaller $c_Q=-0.39155$ merges with another anti-bound mode forming a genuine quasinormal mode (QNM). This mode has a complex frequency $\Omega=\omega+i \Gamma$, which imaginary part $i\Gamma$ corresponds to an exponentially fast time decay of the mode. 

\begin{figure*}
	\includegraphics[width=2\columnwidth]{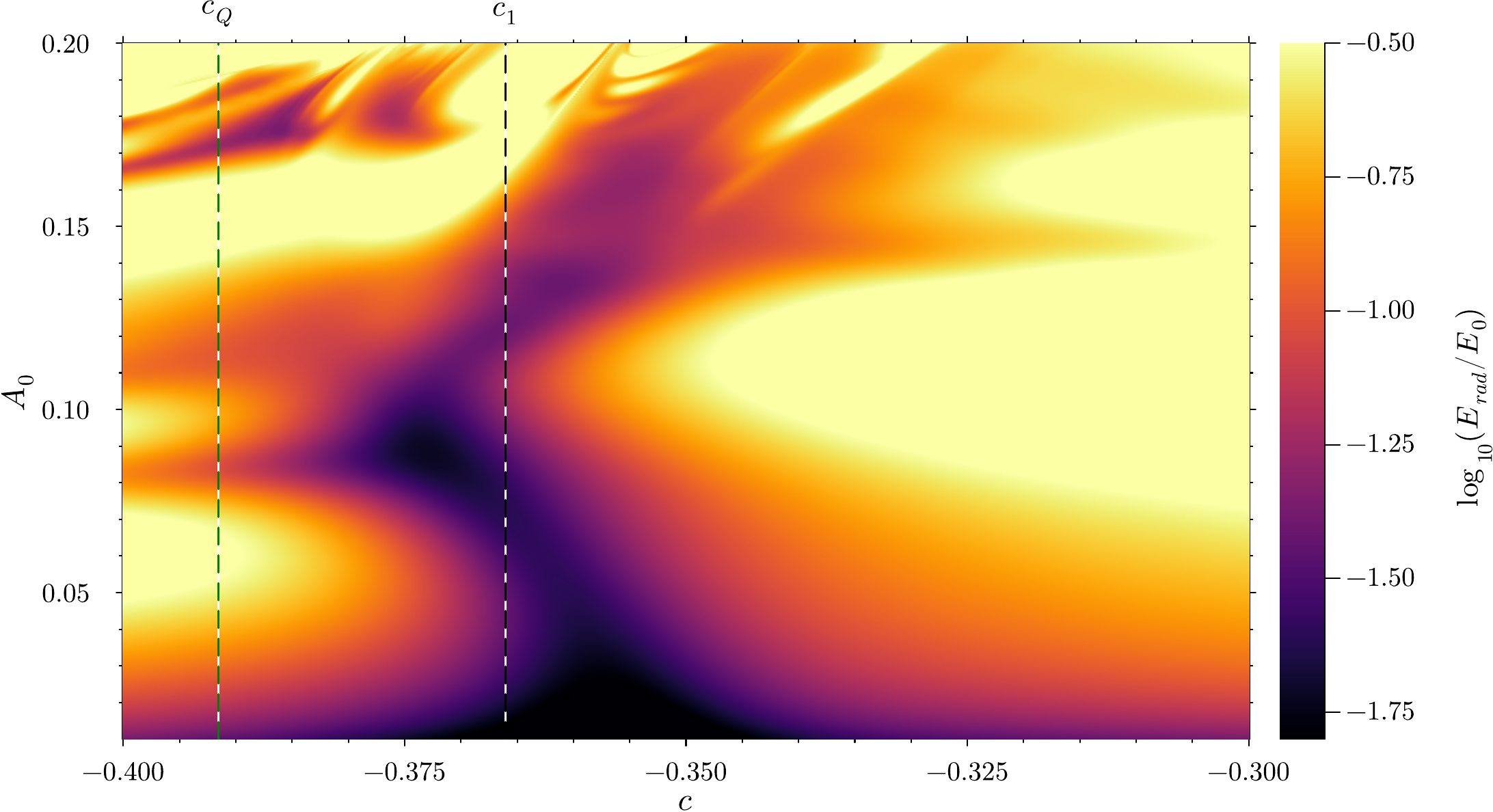} 
	\caption{Energy loss $E_{rad}/E_0$ after $T=300$ for initial pulse with $\beta=0.1$ as a function of the model parameter $c$ and initial amplitude $A_0$. Wobblerons are created near the $c_1$ line in the dark region.} 
    \label{scan-oscillon}
\end{figure*}

To find a wobbleron we use the following initial odd perturbation on top of the CL kink
\begin{equation}
    \phi(x,0)=\phi_K^{(CL)}+A_0 \frac{\tanh(x)}{\cosh(\beta x)},
    \label{initial}
\end{equation}
where $A_0$ and $\beta$ are real parameters related to the amplitude and width of the initial perturbation. This is an odd perturbation. Therefore, it is suitable for the excitation of odd modes. This profile is motivated by eq. (\ref{approx_p1}) which defines the first approximation to the profile of the oscillon in the presence of the kink. Indeed, for $c=c_1$ there is an odd threshold mode $\eta_{th}(x,t)=\tanh (x)\cos (2t)$. 

In Fig. \ref{scan-oscillon} we present the energy loss ratio, $\log_{10} E_{rad}/E_0$, at a late time $T=300$, for $\beta=0.1$. We vary the initial amplitude $A_0$ and the parameter of the CL model $c$ in the vicinity of $c_1$, $c\in [-0.4,-0.3]$. This allows us to analyze four qualitatively different regimes where we have the following modes in the game: (1) a quasinormal mode:  $c<c_{Q}$; (2) an odd anti-bound mode: $c \in (c_{Q},c_1)$; (3) an odd threshold mode: $c=c_1$; (4) an odd bound mode: $c>c_1$. The oscillon (or mode) on the kink is not formed (excited) if the ratio is above -1.

For $c\in (c_Q,c_1)$, an odd wobbleron is formed only if the initial amplitude takes a sufficiently large value. The explanation is straightforward. The mode is again a seed for the wobbleron. However, because it is a non-normalizable anti-bound mode, one needs appropriately large non-linearity, that is, large initial amplitude, to localize it. In this case, there is no oscillon with arbitrary small amplitude. The minimal amplitude tends to zero as the anti-bound mode tends to the threshold mode $c\to c_1$. Now, the wobbleron appears for any value of the initial amplitude. This is because arbitrarily small non-linearities localize the threshold mode and push its frequency below the mass threshold. 

Finally, if the mode of the kink is just the usual discrete normal mode, $c>c_1$, then the initial odd perturbation excites the odd bound mode. 

\begin{figure*}
	\centering
\includegraphics[width=\columnwidth]{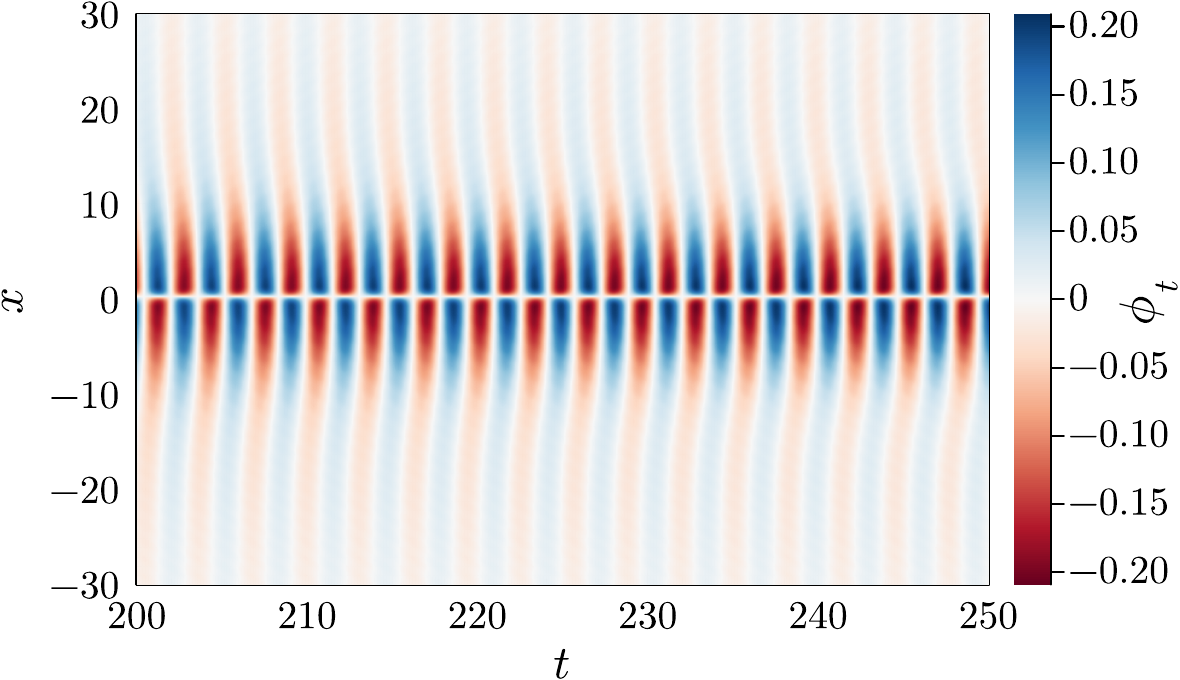}
\includegraphics[width=\columnwidth]{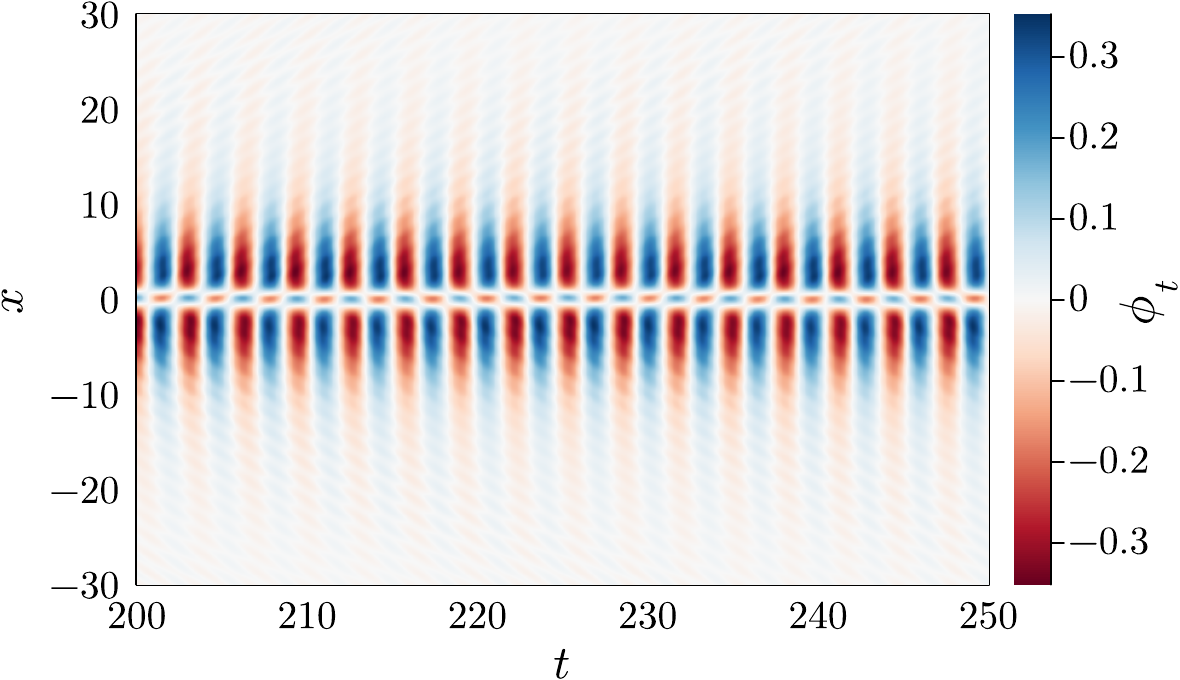}
	\caption{Left: Odd oscillon in CL model with $c=c_1$ and $A_0=0.075$, $\beta=0.1$. Right: Even oscillon in CL model with $c=0$ ($\phi^4$ model) and $A_0=0.1$. To avoid showing the background kink we plot $\phi_t(x,t)$.} 
    \label{CL-oscillon}
\end{figure*}

In Fig. \ref{CL-oscillon}, left panel,  we show an example of the oscillon on the kink, generated from the threshold kink mode, $c=c_1$. 
The actual profile of the odd wobbleron agrees very well with the leading analytic approximation (\ref{approx_p1}) (see Appendix \ref{app:profile}). As in the in-vacuum case, generic initial data often generate the wobbleron in an excited state, which means amplitude modulation. This solution can be interpreted as a bound state of two oscillons on top of the kink. For sufficiently large modulations, we clearly observed two oscillons bouncing around the kink (see Appendix \ref{app:excited}). 

We also analyze the case with an even wobbleron. For that we take $c=c_2$, which is the $\phi^4$ model. Here, the relevant initial data based on the corresponding threshold mode, $\eta_{th}(x,t) = (1-3\tanh^2(x)) \cos (2 t)$, is 
\begin{equation}
    \phi(x,0)=\phi_K^{(c=0)}(x-\delta)+\frac{A_0}{2}\frac{1-3\tanh^2(x)}{\cosh(A_0 x)}, 
\end{equation}
The new parameter $\delta$ has to be added, as even initial data also excite the even zero mode of the kink. Thus, generically, the even wobbleron splits into the kink and a free in-vacuum oscillon. Changing $\delta$.  we can temporarily stabilize the kink. In Fig. \ref{CL-oscillon}, right panel, we show an example of such an even wobbleron

\section{Summary}
In this work, we presented a mechanism that explains the origin of oscillons. We showed that oscillons and internal modes are not independent phenomena. Oscillons emerge from the linear resonances, in particular, non-normalizable threshold and anti-bound modes, which are localized due to the nonlinearity of the model. The condition that guarantees the localization of the mode and therefore the formation of the oscillon is an integral generalization of the in-vacuum condition \cite{Fodor:2008es}. 

An important consequence of this mechanism of the appearance of oscillon is the existence of {\it wobblerons}, i.e., {\it nonlinear excitations of solitons}, which are oscillon-soliton bound states. In particular, odd oscillons can form stable bound states. This is because there are no odd threshold modes in vacuum. Even oscillons have a tendency to excite the zero mode of the kink and eventually split from it. 

Furthermore, the formation of the wobbleron explains the dynamics of the excited kinks if they pass the spectral wall, i.e., a phenomenon that occurs when a normal mode hosted by a soliton hits, during the evolution, the mass threshold \cite{Adam:2019xuc} and \cite{Adam:2022bus, Alonso-Izquierdo:2024fpw, Krusch:2026bhh}. Now, the soliton can stop and form a stationary solution, beyond the point where the bound mode transmutes to the threshold mode. This stationary solution is just a soliton-oscillon bound state; see Appendix \ref{app:sw} for an extended discussion. 

There are no reasons why, as presented here, wobblerons need to be a (1+1) dimensional real scalar field phenomenon only. On the contrary, we expect that higher dimensional solitons as, for example, vortices can form bound states with oscillons. In fact, depending on the coupling constant, vortices host various normal modes. This concerns vortices in the Abelian Higgs model \cite{Alonso-Izquierdo:2024bzy, Alonso-Izquierdo:2025xgu} and in other theories \cite{Gavrea:2026adu}. As the coupling changes, many of them hit the mass threshold and then transmute to resonances, for example to Feshbach resonances \cite{Bachmaier:2025igf}. These modes can give rise to nonlinear oscillon-vortex bound states. Also, the BPS limit can be interesting from this point of view. Here, the mode structure varies as we move in the moduli space in the fixed topological sector (similarly to the BPS-impurity case), that is, as we vary the geometry of the BPS solution \cite{Alonso-Izquierdo:2023cua,Alonso-Izquierdo:2025suz, AlonsoIzquierdo:2026mub}. Again, for a particular geometry, some of the modes can become threshold modes which, if we further change the BPS solution, transform to resonances that can trigger the oscillon inside the BPS vortex. Wobblerons can also exist in complex field models and take part in $Q$-ball dynamics \cite{Copeland:2014qra,Alonso-Izquierdo:2025iet, Martinez:2025ana, Blaschke:2025anm}. 

In this work we focused on (1+1) dimensional oscillons. In higher dimensions, the structure of the threshold and anti-bound resonances is more involved. E.g., in two spatial dimensions the transition from exponentially divergent anti-bound mode to exponentially localized bound mode goes via modes with logarithmic tails. The qualitative difference of the threshold modes seems to be reflected in different properties of higher dimensional oscillons. Nevertheless, the framework introduced in the current work can still be applied to such oscillons.

\section*{Acknowledgements}
FB have been supported by the grant no. SGS/24/2024 Astrophysical processes in strong
gravitational and electromagnetic fields of compact object.
AW and KS have been supported in part by Spanish Ministerio de Ciencia e Innovacion (MCIN) with funding from the grant PID2023-148409NB-I00 MTM.
KS acknowledges financial support from the Polish National Science
Centre (Grant No. NCN 2021/43/D/ST2/01122).

\newpage 

\begin{center}
    \textbf{\large Supplemental material}
\end{center}

\appendix
\section{The envelope formula}
\label{app:envelope}

Here, we will find the localized function. Acting with Hamiltonian $\mathcal{H}$ on (\ref{p1}) and expanding derivatives
we find
\begin{equation}
\mathcal{H} p_1 =-\kappa^2 \eta(x)F
-2\epsilon \eta'(x)F'
- \epsilon^2 \eta(x)F''.
\end{equation}
Now, equation (\ref{full}) becomes
\begin{equation}
2\epsilon  \eta' F'
+\epsilon^2 \eta F''
+
\epsilon^2\Bigl( \tilde{\omega}_2 \eta F +\lambda(x) \eta^3 F^3\Bigr)
=0,
\end{equation}
where $\tilde{\omega}_2=\omega_2- \kappa^2/\epsilon^2$.
At order $\epsilon$, the first term vanishes upon projection
\begin{equation}
\int_{-L}^{L}  \eta(x) \eta'(x)\,dx = 0,
\end{equation}
where we introduce a cut-off due to the non-normalizability of the mode. Equivalently, the cut-off arises due to the localization enforced by the envelope.
At order $\epsilon^2$, multiplying by $\eta(x)$ and integrating, we obtain the solvability condition
\begin{align}
F'' \int_{-L}^{L}  \eta^2(x)\,dx &+
\tilde{\omega}_2  F \int_{-L}^{L}  \eta^2(x)\,dx
+  \nonumber \\
+ F^3 \int_{-L}^{L}  \lambda(x) \eta^4(x)\,dx
=0,
\end{align}
Defining the coefficients
\begin{equation}
C_2 = \int_{-L}^{L}  \eta^2(x)\,dx,
\qquad
C_4 = \int_{-L}^{L}  \lambda(x)  \eta^4(x) \,dx,
\end{equation}
we arrive at the effective envelope equation
\begin{equation}
F''(X) =-\tilde{\omega}_2 F(X) - g\,F^3(X),
\qquad
g = \frac{C_4}{C_2}.
\end{equation}
This equation admits localized solutions of the form
\begin{equation}
F(X) = \sqrt{-\frac{2\tilde{\omega}_2}{g}}\mathrm{sech}\bigl(\sqrt{-\tilde{\omega}_2}\,X\bigr),
\end{equation}
provided $\tilde{\omega}_2<0$ and $g>0$. The last condition can be equivalently written as
\begin{equation}
 \int_{-\infty}^{\infty}
 \lambda(x)\, \eta^4(x)  \,F^4(\epsilon x)\,dx >0. \label{omega2}
\end{equation}
This is the in-medium generalization of the well-known $\lambda >0$ condition for the existence of small oscillons derived in \cite{Fodor:2008es}. 

Consequently, the oscillon profile takes the asymptotic form
\begin{equation}
p_1(x) \sim \eta(x)\,\mathrm{sech}(\sqrt{-\tilde{\omega}_2}\,\epsilon x),
\end{equation}
where $\tilde{\omega}_2=\omega_2- \kappa^2/\epsilon^2$.
This proves formula (\ref{approx_p1}). 


Note that we have not yet specified entirely boundary conditions for the solution of equation (\ref{full}). Normalizability implies that $p_1$ should vanish at infinities, but we need one more condition to uniquely define the solution. Equation (\ref{full}) possesses scaling symmetry, $p_1(x)\to\alpha p_1(x)$, $\omega_2\to\alpha^2\omega_2$ and $\epsilon\to \epsilon/\alpha$ which basically transforms a solution into a solution with a different amplitude and for a different $\epsilon$. We can use this symmetry to fix $\tilde \omega_2=-1$. This leads to our final expression (\ref{approx_p1}).

\section{Wobbler in Sine-Gordon}
\label{app:sG}
\begin{figure}
	\centering
    \includegraphics[width=\columnwidth]{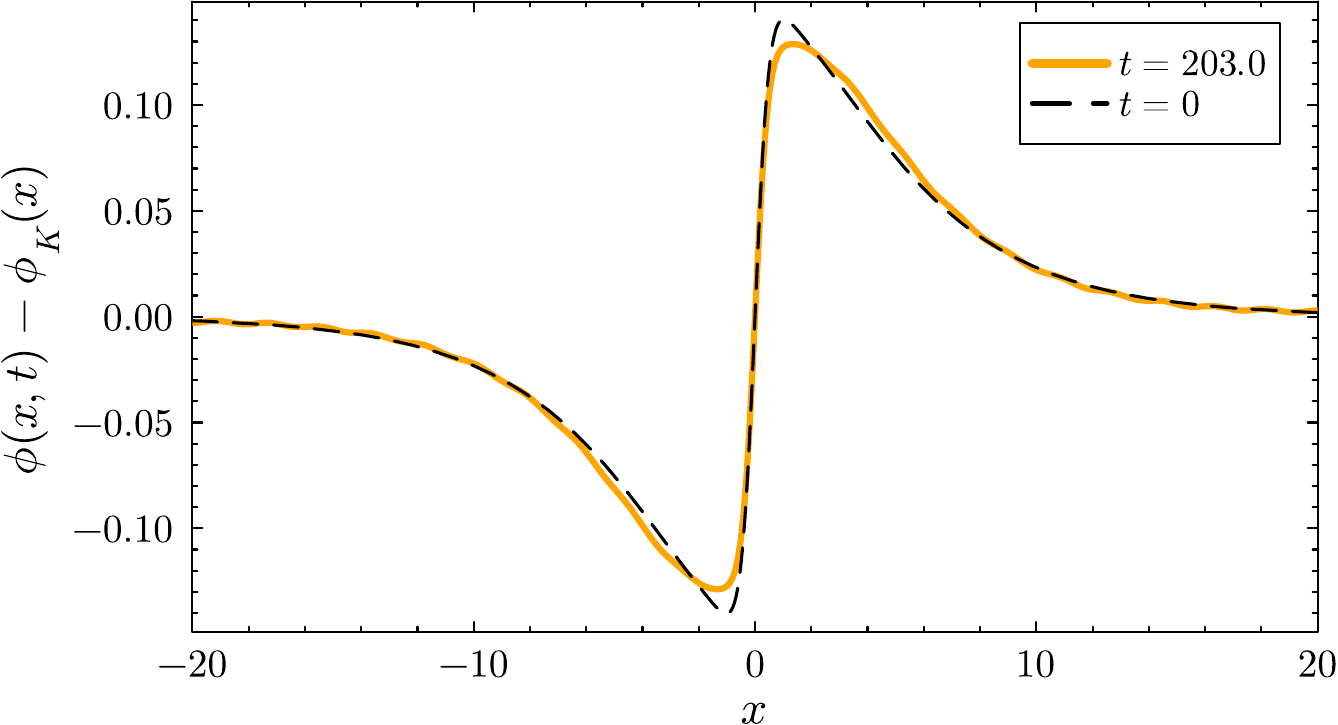}

    \includegraphics[width=\columnwidth]{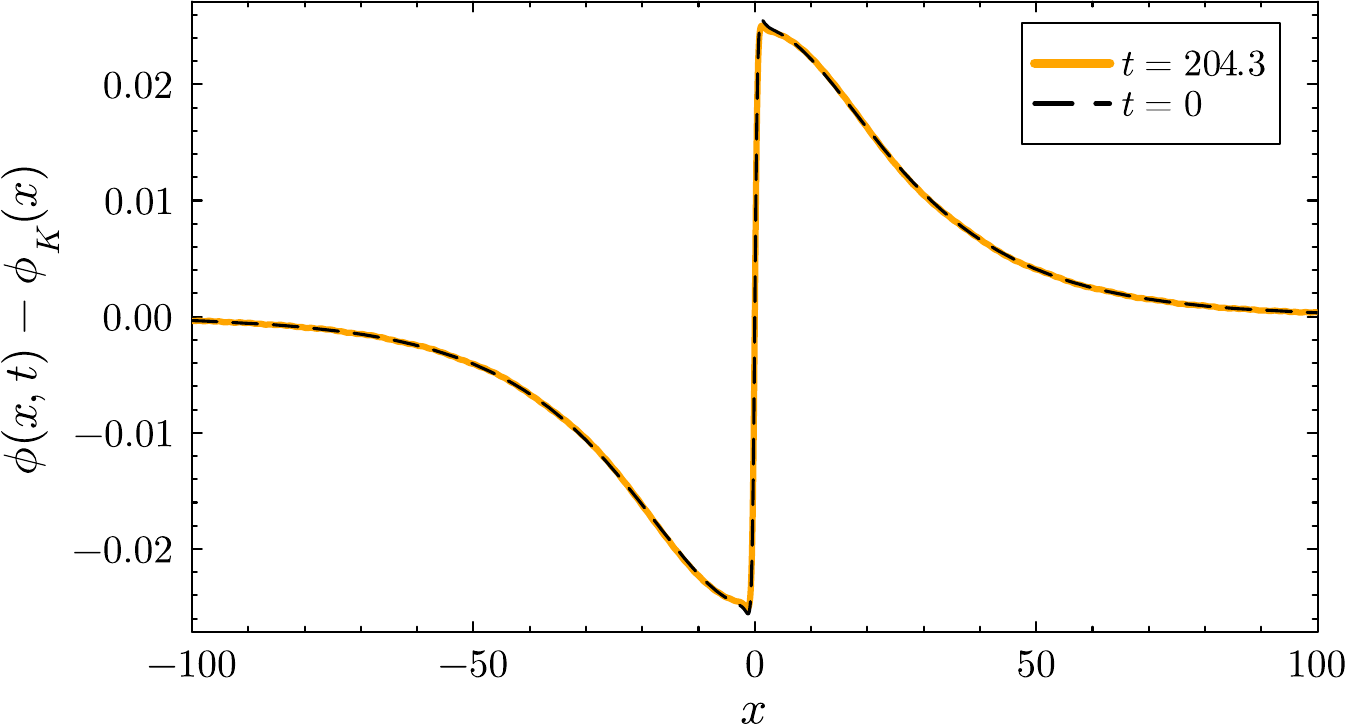}
	\caption{Comparison of the profile of the initial perturbation and at the first maximum after $t=200$ for $\beta=0.25$, $A=0.14217$ (uppeer panel) and $\beta=0.05$, $A=0.025000$ (lower panel) and $c=c_1$. Optimal amplitude was found by minimizing energy flux for fixed $\beta$.} 
    \label{Profiles}
\end{figure}

Here we consider the sG model
\begin{equation}
    U_{sG} = 1-\cos \phi
\end{equation}
for which the kink reads
\begin{equation}
    \phi_{K}^{(sG)} = 4\arctan e^{x}.
\end{equation}
Then, the functions entering into the oscillon-profile equation are
\begin{equation}
    V(x)=-\frac{2}{\cosh^2 x}
\end{equation}
and
\begin{equation}
    g_2(x)=-\frac{\tanh x} \cosh {x}, \;\;\;\; g_3(x)=-\frac{1}{6}+\frac{1}{3\cosh^2 x}. 
\end{equation}
The sG kink has the obvious zero mode $\eta_0$ and threshold mode $\eta_{th}$
\begin{equation}
    \eta_0(x,t)=\frac{1}{\cosh x}, \;\;\; \eta_{th}(x,t)=\tanh x \cos t.
\end{equation}
Here, the mass threshold is $m=1$. 

Using the integrability of the theory and, in particular, the existence of the exact kink-breather solution, the so-called {\it wobbler}, we can see how the oscillon localized on the kink goes to the threshold mode. Such a three-soliton time-periodic and spatially antisymmetric kink-breather solution can be written as

\begin{equation}
\phi(x,t)=\phi_K(x)+\psi(x,t),
\end{equation}
where
\begin{widetext}
\begin{equation}
    \psi(x,t)=
4\,\arctan\!\left(
\frac{2\epsilon(1+\epsilon)}{\omega^2}
\frac{
\cos(\omega t)\,\sinh x - \sinh(\epsilon x)
}{
\left(\cos(\omega t) + \cosh\!\big((1+\epsilon)x\big)\right)^2
+
\frac{(1+\epsilon)^2}{\omega^2}
\Big[
\big(-\cos(\omega t)\,\sinh x + \sinh(\epsilon x)\big)^2
+
\cosh^2 x\,\sin^2(\omega t)
\Big]
}
\right)
\end{equation}
\end{widetext}
and $\epsilon=\sqrt{1-\omega^2}$. Importantly, expanding the oscillating part at $\epsilon \to 0$ we find
\begin{equation}
    \psi(x,t)= 4\epsilon \sin (\omega t) \frac{\tanh x}{\cosh (\epsilon x)}. 
\end{equation}
As predicted by our theory, this is exactly the threshold mode $\eta_{th}=\tanh x \sin t$ of the sG kink localized by the envelope $F(\epsilon x)=\sech (\epsilon x)$. Its frequency is reduced $\omega <1$.

\section{Comparison of the wobbleron with the profile of the localized threshold mode}
\label{app:profile}

In Fig. \ref{Profiles} we compare the profile of the odd wobbleron found in full numerics with the leading analytic approximation (\ref{approx_p1}). Here we consider the CL model with $c=c_1$, i.e., where an odd threshold mode exists. We found an excellent agreement. In these examples, the amplitude $A_0$ is chosen to minimize the initial radiation.

\section{Excited wobblerons}
\label{app:excited}
\begin{figure}
    \centering
    \includegraphics[width=\linewidth]{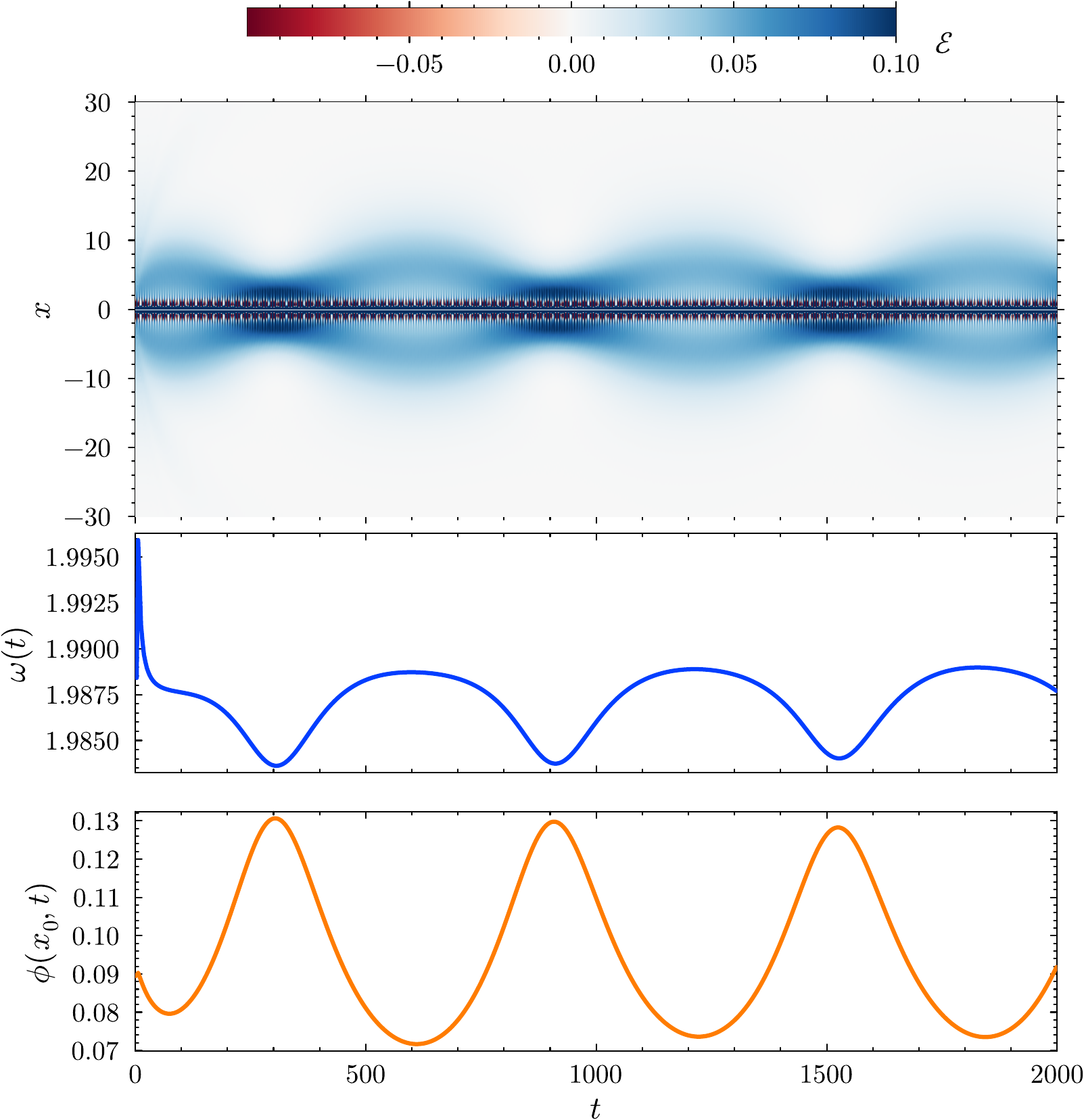}
    \caption{Example of an excited wobbleron in CL model with $c=-0.373$, which is a bound state of kink and two oscillons. Here, $x_0=2$ and  $A_0=0.09$ and $\beta=0.1$.}
    \label{fig:2-osc-kink}
\end{figure}

    \begin{figure}
    \includegraphics[width=\linewidth]{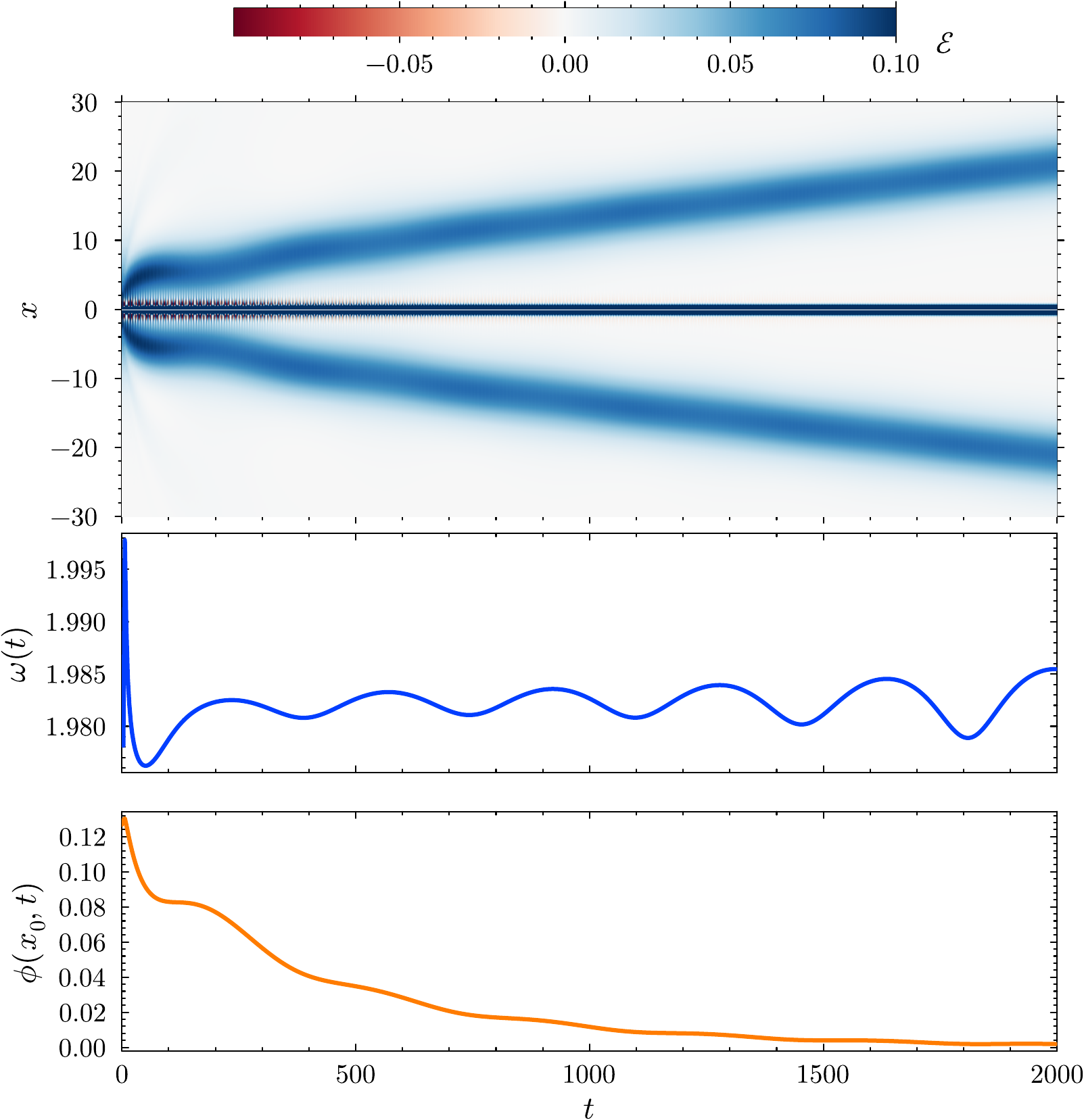}
    \caption{Example of the decay of an excited wobbleron in CL model with $c_1=-0.386$, where excited wobbleron splits into  two oscillons leave the kink. Here, $x_0=2$ and $A_0=0.130753$ 
    and $\beta=0.15$.}
    \label{fig:2-osc-kink-free}
\end{figure}

\begin{figure}
	\centering
    \includegraphics[width=\columnwidth]{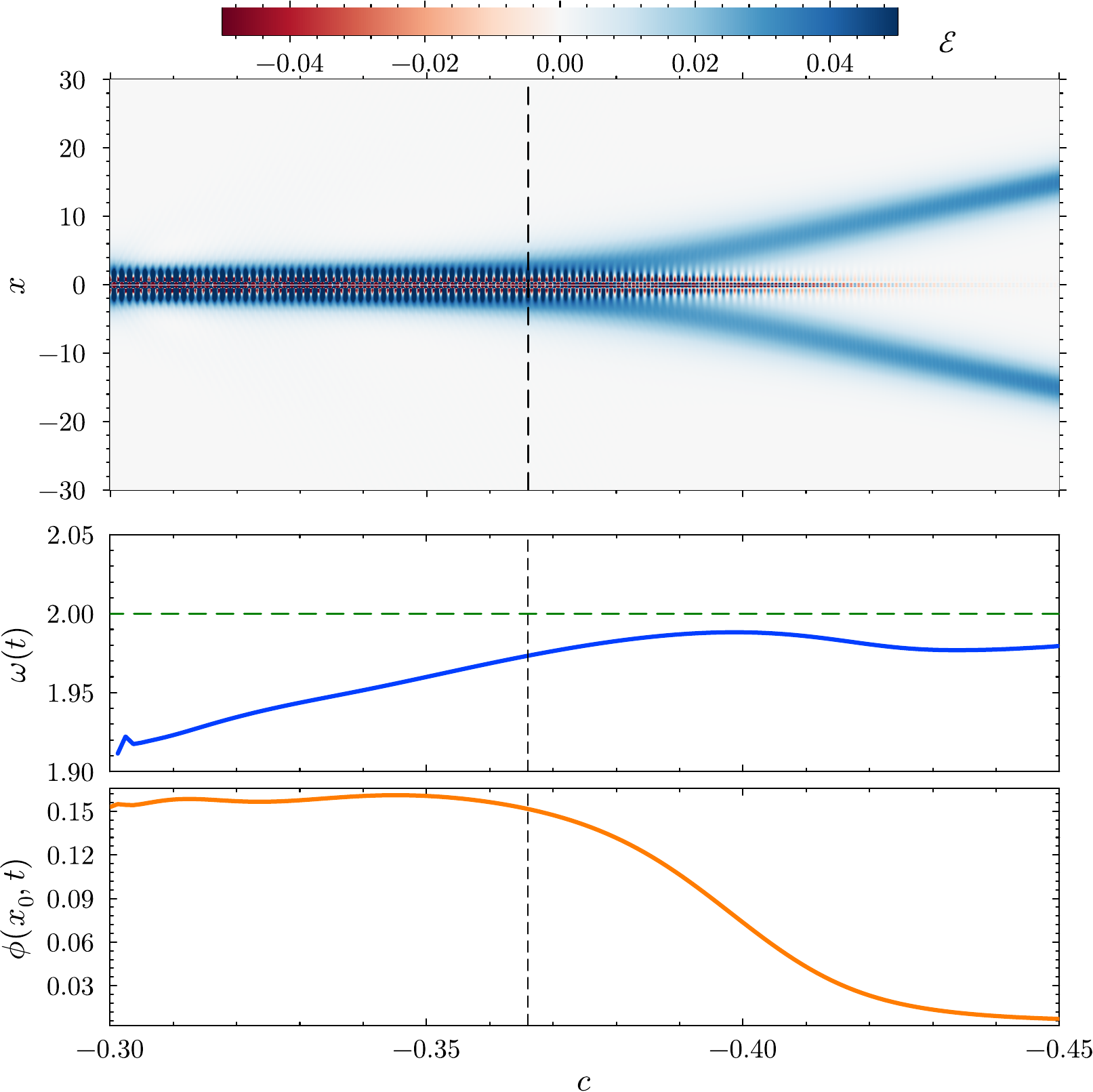}
	\caption{Wobbleron parameters during adiabatic evolution $c(t)=c_i+(c_f-c_i)t/T$ starting with  a bound mode for $c_i=-0.3$. $c_f=-0.45$, $T=4000$ with initial amplitude $A_i=0.11$}
    \label{fig:adiabatic-CL}
\end{figure}

As in the in-vacuum case, generic initial data often generate the wobbleron in an excited state, which means the amplitude modulation. This solution can be interpreted as a bound state of two oscillons on top of the kink. For sufficiently large  modulations, we clearly observed two oscillons bouncing around the kink, see Fig. \ref{fig:2-osc-kink}. Changing the initial data, e.g., the amplitude $A_0$ of the initial pulse, we find a smooth transition to the scenario in which the excited wobbleron breaks up. These two oscillons leave the kink and become in-vacuum oscillons traveling with constant velocity, see Fig. \ref{fig:2-osc-kink-free}. 

It is worth to mention that these two last examples ($c<c_1$) concern the case where the wobbleron is seeded by the anti-bound mode rather than the threshold mode.   

Finally, in Fig. \ref{fig:adiabatic-CL} we present an adiabatic generation of the wobbleron starting from the kink with the bound mode excited. In this approach, we begin with the CL model in the regime where it supports the $n=1$ odd mode. We chose $c_i=-0.3$. Using the initial data (\ref{initial}) we excite this mode. Then, we slowly change the parameter of the CL model
\begin{equation}
    c(t)=c_i + \frac{t}{T} (c_f-c_i),
\end{equation}
where $c_f=-0.49$ and $T=4000$. Initially, as $c$ decreases, the solution settles on the corresponding excited kink. However, at certain time when $c=c_1$, the bound mode transmutes into the anti-bound mode and the wobbleron, i.e., oscillon-kink bound state appears. As $c$ further decreases, the anti-bound mode changes to a quasi-normal mode. Then, the bound state breaks up and we find kink with free oscillons in the final state.

\section{Spectral walls}
\label{app:sw}

Until now, the appearance of the threshold and antibound resonances was governed by the value of the parameter in the CL model. Here, we will focus on a theory in which the existence and property of the mode changes in a {\it dynamical way}. This means that a normal bound mode can dynamically, during the evolution of the system, change into the threshold resonance and then transform to the antibound and genuine quasinormal mode. In fact, a dynamical change of the mode structure always occurs in solitonic systems whenever one soliton feels the presence of another soliton, see e.g. the case of two critically coupled vortices in the Abelian Higgs theory \cite{Alonso-Izquierdo:2023cua}. This can have a significant impact on the dynamics of solitons, especially in BPS models \cite{Krusch:2024vuy, AlonsoIzquierdo:2024nbn, Alonso-Izquierdo:2025suz, AlonsoIzquierdo:2026mub, Krusch:2026bhh}. However, the most striking manifestation of the flow of the spectral structure is the {\it spectral wall} \cite{Adam:2019xuc}. 

Spectral wall is a phenomenon that occurs when a normal mode hosted by a soliton hits, during the evolution, the mass threshold. It acts as a barrier in the motion of the soliton. Specifically, for a fine-tuned (critical) amplitude $A_{cr}$ of the mode, the initially moving soliton stops and forms a long-lived stationary state, in a spatial point that corresponds to the configuration for which the mode becomes a threshold mode. For smaller amplitudes, $A<A_{cr}$, the soliton passes through this point, although its motion becomes more and more deformed as $A\to A_{cr}$. For $A>A_{cr}$ the soliton is strongly repelled from the point where the spectral wall is located, which results in backscattering. 

This phenomenon was originally discovered in a BPS-kink dynamics in (1+1) dimensions \cite{Adam:2019xuc}, but it exists in weakly non-BPS models \cite{Adam:2022bus, Martinez:2025ana} or in higher dimensions, see, e.g., vortices in the Abelian Higgs model at the critical coupling \cite{Alonso-Izquierdo:2024fpw, Krusch:2026bhh}. 

The presented unification of the modes and oscillons sheds new light on the problem of the spectral wall. Especially, it can explain dynamics of a soliton beyond the spectral wall, i.e., where the threshold mode continues as an anti-bound mode. 

The spectral wall is a linear response phenomenon in which the repulsive soliton-barrier force and the position of the spectral wall can be found in the linear perturbation theory. As we know, the threshold mode does not disappear but rather smoothly passes to an anti-bound mode. Therefore, for excited solitons that pass the spectral wall, e.g., for those with sufficiently large velocity, or not to large amplitude, the energy stored in the initially excited bound mode goes to the anti-bound mode. If the amplitude of the mode is large enough to trigger non-linearities of the model, but sufficiently small to allow the soliton to go through the spectral wall (no backscattering), the excited anti-bound mode may transmute into the ocillon located on the soliton. Then the woblleron is formed. As described above, the localization of the resonance is a nonlinear perturbation phenomenon. 
\begin{figure}
	\centering
\includegraphics[width=\columnwidth]{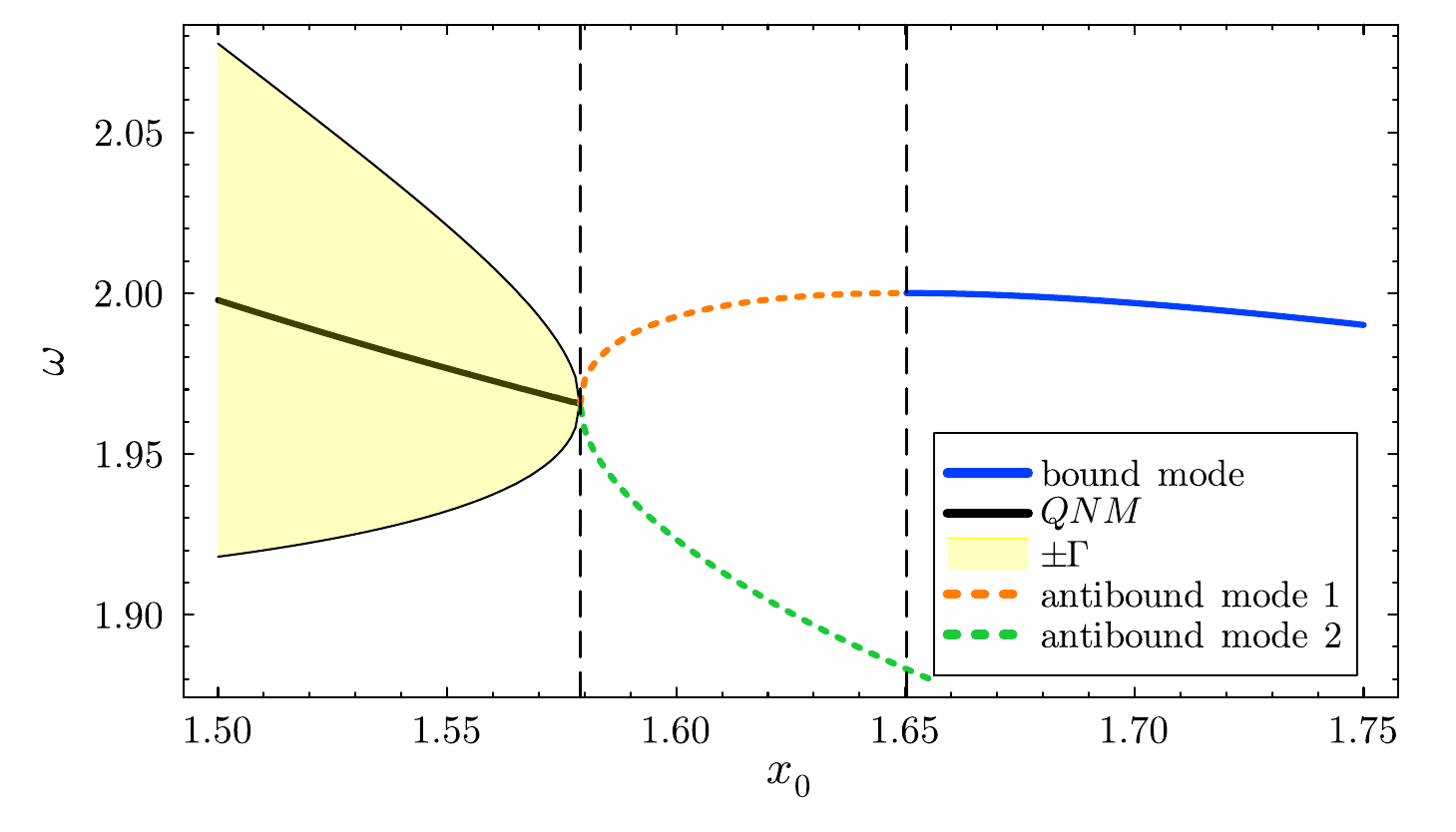}
	\caption{Structure of the linear modes of the antikink in the BPS impurity deformed $\phi^4$ model with impurity given by (\ref{imp}).} 
    \label{BPS-modes}
\end{figure}

As an example, we will consider the BPS-impurity version of the $\phi^4$ model \cite{Adam:2018tnv, Adam:2019xuc}
\begin{equation}
 L=\int_{-\infty}^\infty  dx\left[ \frac{1}{2} \phi_t^2 - \frac{1}{2} \left( \phi_x + (1-\phi^2) + \sqrt{2} \sigma \right)^2 \right],
 \end{equation}
 where for concreteness the impurity (non-dynamical background field $\sigma(x)$) is 
 \begin{equation}
     \sigma(x)=\frac{3}{\cosh^2 x}. \label{imp}
 \end{equation}
 This theory has a one-parameter family of energetically degenerate static  solutions $\phi_{AK}(x;x_0)$ that solve the Bogomolny equation
 \begin{equation}
  \phi_x + (1-\phi^2) + \sqrt{2} \sigma  =0
 \end{equation}
and describe an antikink at any distance $x_0$ from the impurity (located at the origin). For $x_0\to \infty$ the solution becomes the $\phi^4$ antikink. The position of the BPS antikink corresponds to the zero of the field $\phi_{AK}(x=x_0;x_0)=0$, see \cite{Adam:2018tnv, Adam:2019xuc} for details. 

Although energetically equivalent, the BPS anti-kinks have the $x_0$-dependent structure of linear modes, see Fig. \ref{BPS-modes}. Importantly, there is an odd normal bound mode for $x_0>x_{sw}$ (blue curve), which asymptotically, as $x_0\to \infty$ tends to the shape mode of the $\phi^4$ kink. At $x_{sw}\approx 1.65$ we have a spectral wall where the mode becomes the threshold mode. For $x<x_{sw}$ it changes to an anti-bound mode, and then finally merges with another anti-bound mode and forms a quazinormal mode with a complex frequency $\Omega=\omega+i\Gamma$, see also \cite{Evslin:2022xmp}. 

\begin{figure}
	\centering
\includegraphics[width=\columnwidth]{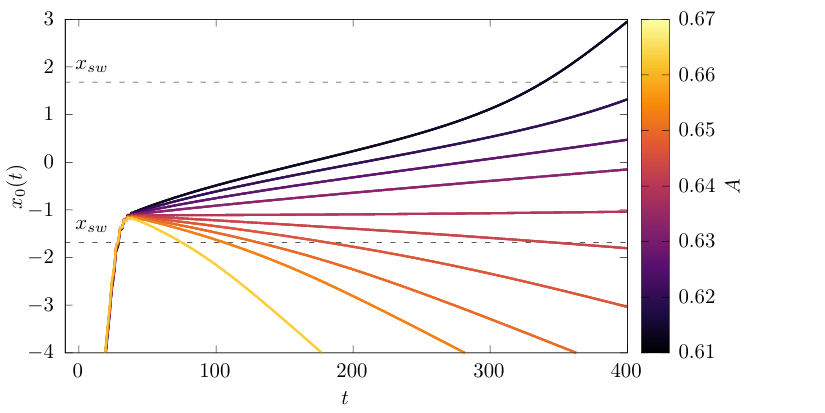}
	\caption{Evolution of the excited BPS antikink with $v=0.25$ and initial amplitude $A$. For $A=A_{cr}=$ we see formation of a stationary solution beyond the spectral wall $x_{sw}$. } 
    \label{SW}
	\centering
\includegraphics[width=\columnwidth]{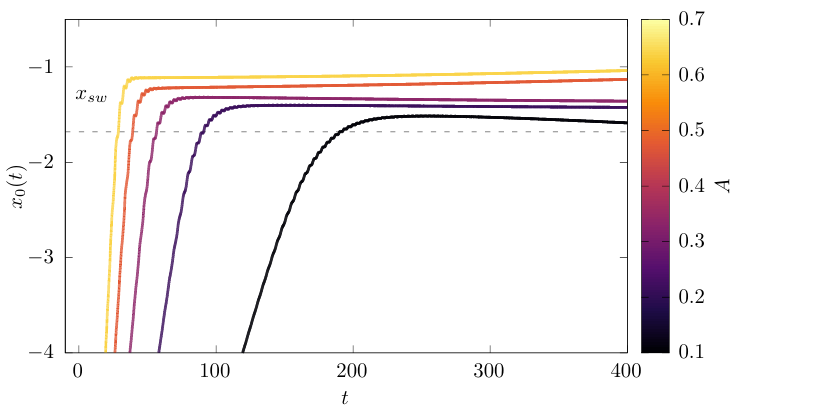}
	\caption{Formation of the stationary solutions beyond the position of the spectral wall. Here, $v\in[0.05,0.25]$, $\Delta v=0.05$.} 
    \label{SW-v}
\end{figure}

The lowest energy dynamics, i.e., a boosted antikink, is just evolution through the available BPS configuration. This is nothing but geodesic motion along the corresponding moduli space of the BPS solutions. The only mode excited is the zero mode (kinetic degrees of freedom) that moves one BPS solution into another.

Now we excite the massive normal mode. For a small velocity $v$ of the incoming antikink we are still in the linear perturbation theory regime. As the frequency of the mode increases, as the antikink approaches the impurity, the excited mode introduces a repulsive force. Then, at $x=x_{sw}$ we see a spectral wall. The critically excited soliton forms a long-lived stationary state at $x=x_{sw}$, as predicted from the spectral structure \cite{Adam:2019xuc}. The critical amplitude $A_{cr}$ is approximately a linear function of $v$.

For increasing $v$ we need a large amplitude $A$ to find the stationary solution. At some point, we enter the nonlinear perturbation regime. In Fig. \ref{SW} we show the evolution of the antikink for $v=0.25$ and several values of the initial amplitude $A$. We plot the zero of the field identified with the position of the soliton. We clearly see the formation of a stationary state at $x<x_{sw}$. Of course, for $x<x_{sw}$ the initially excited mode is an anti-bound mode. However, as the amplitude is large enough, the nonlinearity transforms it to an oscillon on top of the antikink. 

In Fig. \ref{SW-v} we present the formation of the stationary state for different values of the initial velocity $v$ found for pertinent critical amplitudes $A_{cr}$. As $v$ decreases, we are back in the linear perturbation regime and the position of the stationary solution approaches the location of the spectral wall $x=x_{sw}$.

\bibliography{modes_and_oscillons}

\end{document}